\documentclass[%
 reprint,
superscriptaddress,
 amsmath,amssymb,
 aps,
]{revtex4-2}
\usepackage{graphicx}
\usepackage{dcolumn}
\usepackage{bm}

\begin{document}

\title{Laser-induced Metastable Magnetization in Altermagnets via Ultrafast Asymmetric Spin Dynamics}

\author{Zhaobo Zhou}
\affiliation{Faculty of Science, Charles University, Prague 12843, Czech Republic
}

\author{Sangeeta Sharma}
 \email{sharma@mbi-berlin.de}
\affiliation{Max-Born-Institute for Non-linear Optics and Short Pulse Spectroscopy, Max-Born Strasse 2A, 12489 Berlin, Germany
}
\affiliation{Institute for Theoretical Solid-State Physics, Freie Universität Berlin, Arnimallee 14, 14195 Berlin, Germany
}

\author{John Kay Dewhurst}
\affiliation{Max-Planck-Institut fur Mikrostrukturphysik, Weinberg 2, D-06120 Halle, Germany
}

\author{Junjie He}
 \email{junjie.he.phy@gmail.com} 
\affiliation{Faculty of Science, Charles University, Prague 12843, Czech Republic
}

\date{\today}
\begin{abstract}
Laser pulses are known to induce symmetric demagnetization: equal loss of magnetic moments in the identical sublattices of antiferromagnets and ferromagnets at ultrashort timescales. Using time-dependent density functional theory, we show that linearly polarized laser pulses can drive asymmetric demagnetization between otherwise identical sublattices in the $d$-wave compensated altermagnet (AM) RuO$_2$, resulting in a \textit{photo-induced ferrimagnetic state} with a strong net magnetization of $\sim$0.2 $\mu_B$ per unit cell. The sign and magnitude of this metastable magnetization are highly controllable by laser polarization. We identify polarization-selective asymmetric optical intersite spin transfer (a-OISTR) as the primary mechanism generating the net moment, followed by asymmetric spin flips (a-SF) that further amplifies it. Both effects originate from the characteristic nodal spin band topology of \textit{d}-wave AMs. Moreover, we demonstrate that this laser-induced magnetization is universal across various $d$-wave AMs, including experimentally confirmed KV$_2$Se$_2$O and RbV$_2$Te$_2$O. We uncover a robust route to light-controlled magnetization in AMs on ultrafast timescales.
\end{abstract}

\maketitle

\textit{Introduction}--Magnetism has long played a crucial role in both fundamental research and technological applications. The recent discovery of altermagnet (AM), a novel class of magnetism, breaks the long-standing paradigm that has defined magnetic materials for the past few decades \cite{1,2,song}, and has been named a big breakthrough of 2024 \cite{3}. Unlike conventional ferromagnet (FM) and antiferromagnet (AFM), AMs are characterized by a compensated magnetic structure, where the net magnetization is strictly zero, yet time-reversal symmetry is broken \cite{4,5,6}. These distinct spin properties give rise to alternating and non-relativistic spin splitting in momentum (\textbf{\textit{k}}) space, even in the absence of spin-orbital coupling \cite{7,8}. 

AM has “dual-phase” magnetic behavior, combining the merits of FM and AFM while exhibiting unique properties unparalleled in either of the conventional magnetic classes. These remarkable characteristics position AMs as a promising platform for novel AM-based spintronics applications \cite{9}. For instance, by means of its unique alternating spin splitting, RuO$_2$ has been reported experimentally and theoretically to generate spin-current and current-induced spin torque \cite{Bai1,Bai,bose,gonz,zhang2025,Karube,Anna,shao2023neel,shao2021spin}, facilitating the discovery of exotic quantum phenomena in AM. Besides, their anisotropic \textit{d}/\textit{g}/\textit{i}-wave nodal electronic structures not only lead to direction-dependent transport properties but also could induce an anisotropic response to light.

Despite this, the field of AMs is in its infancy and little is known about the behavior of this interesting class of materials, with multi-component magnetism, under laser pumping. A key theoretical breakthrough in the field of laser-pumped multi-component magnets was the proposal of the optical-induced intersite spin transfer (OISTR) effect proposed by Dewhurst et al. \cite{10}, which showed that optical excitation can coherently redistribute spins between the magnetic sublattices. Notably, OISTR has been experimentally confirmed in various magnetic systems, opening the way for magnetic control towards attosecond timescales \cite{11,ryan2023optically,13,14,15}. Combining these two concepts makes the core of our work.

In this letter, we uncover asymmetric sublattice magnetization dynamics in the fully compensated altermagnet RuO$_2$, giving rise to a metastable photo-induced ferrimagnetic state. This state originates from asymmetric (a) OISTR effect, followed by asymmetric spin flip (a-SF) that further amplifies this asymmetry. Both mechanisms stem directly from the characteristic nodal spin structure of \textit{d}-wave AMs. Notably, this behavior cannot occur in conventional magnets, where identical local environments and an \textit{s}-wave Fermi surface enforce symmetric demagnetization between magnetic sublattices. In AMs, however, a-OISTR and a-SF enable linearly polarized light to drive direction-dependent asymmetric spin dynamics, allowing ultrafast and controllable magnetic states transition. These results provide a previously unrecognized route for manipulating magnetization in AMs on femtosecond timescales.

\textit{Computational Methods}---Our study employs a fully non-collinear spin formulation of real-time time-dependent density functional theory (rt-TDDFT) \cite{runge} to investigate laser pulse-induced ultrafast spin dynamics in AM. This approach incorporates electron interactions through the time-dependent Kohn-Sham (KS) equation, which describes non-interacting electrons evolving in an effective potential. The time-dependent KS equation is given by:

\begin{figure*}
  \centering
  \includegraphics[width=0.7\textwidth]{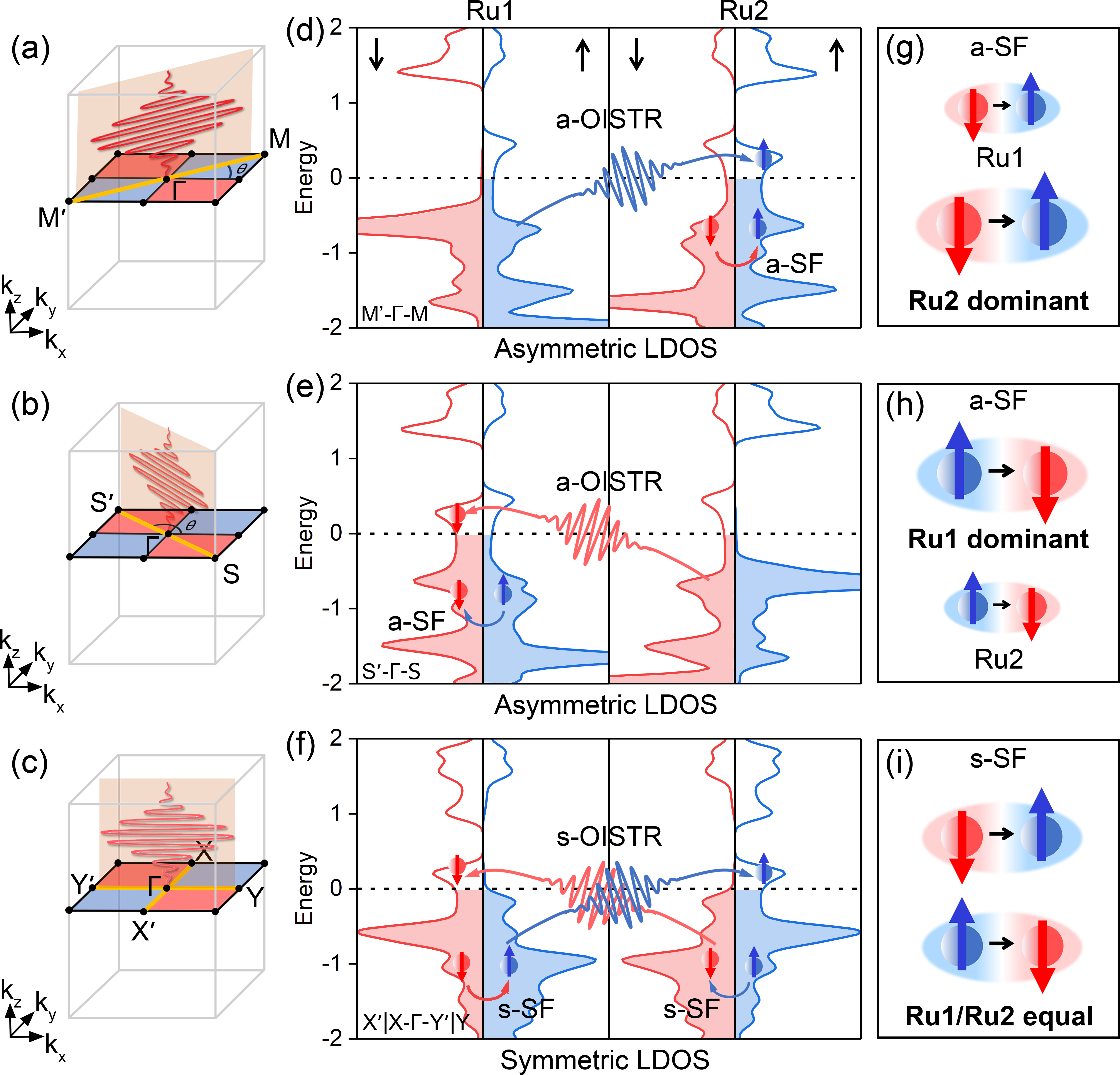}
  \caption{Plorization-dependent \textit{symmetric} (s) and \textit{asymmetric} (a) OISTR and SF mechanism in RuO$_2$. (a-c) 3D Brillouin zone of RuO$_2$. Red and blue regions highlight the alternating symmetry of spin polarization at the nodal plane $k_z$=0. The laser pulses are applied with a various polarization angle $\theta$. Here, the $\hat{\mathbf{e}}_{\theta}$ is set along the spin-polarized M$^{\prime}$--$\Gamma$--M path ($\theta = 45^\circ$), S$^{\prime}$--$\Gamma$--S path ($\theta = 135^\circ$) and spin-degenerate Y$^{\prime}$--$\Gamma$--Y path ($\theta = 0^\circ$), X$^{\prime}$--$\Gamma$--X path ($\theta = 90^\circ$), respectively. (d-f) LDOS of two Ru atoms without SOC along the M$^{\prime}$--$\Gamma$--M,  S$^{\prime}$--$\Gamma$--S and X$^{\prime}$--$\Gamma$--X (Y$^{\prime}$--$\Gamma$--Y) paths. The a-OISTR and a-SF occur between spin-up (blue) and spin-down (red) channels at $\theta = 45^\circ$ and 135$^\circ$ while s-OISTR and s-SF occur at $\theta = 0^\circ/90^\circ$. (g-i) Corresponding schematic of dominant SF from spin-down to spin-up channel for Ru2 at $\theta = 45^\circ$ and dominant SF from spin-up to spin-down channel for Ru2 at $\theta = 135^\circ$. The SF in Ru1 and Ru2 at $\theta = 0^\circ/90^\circ$ are equal.}
  \label{fig:1}
\end{figure*}

\begin{figure*}
  \includegraphics[width=0.7\textwidth]{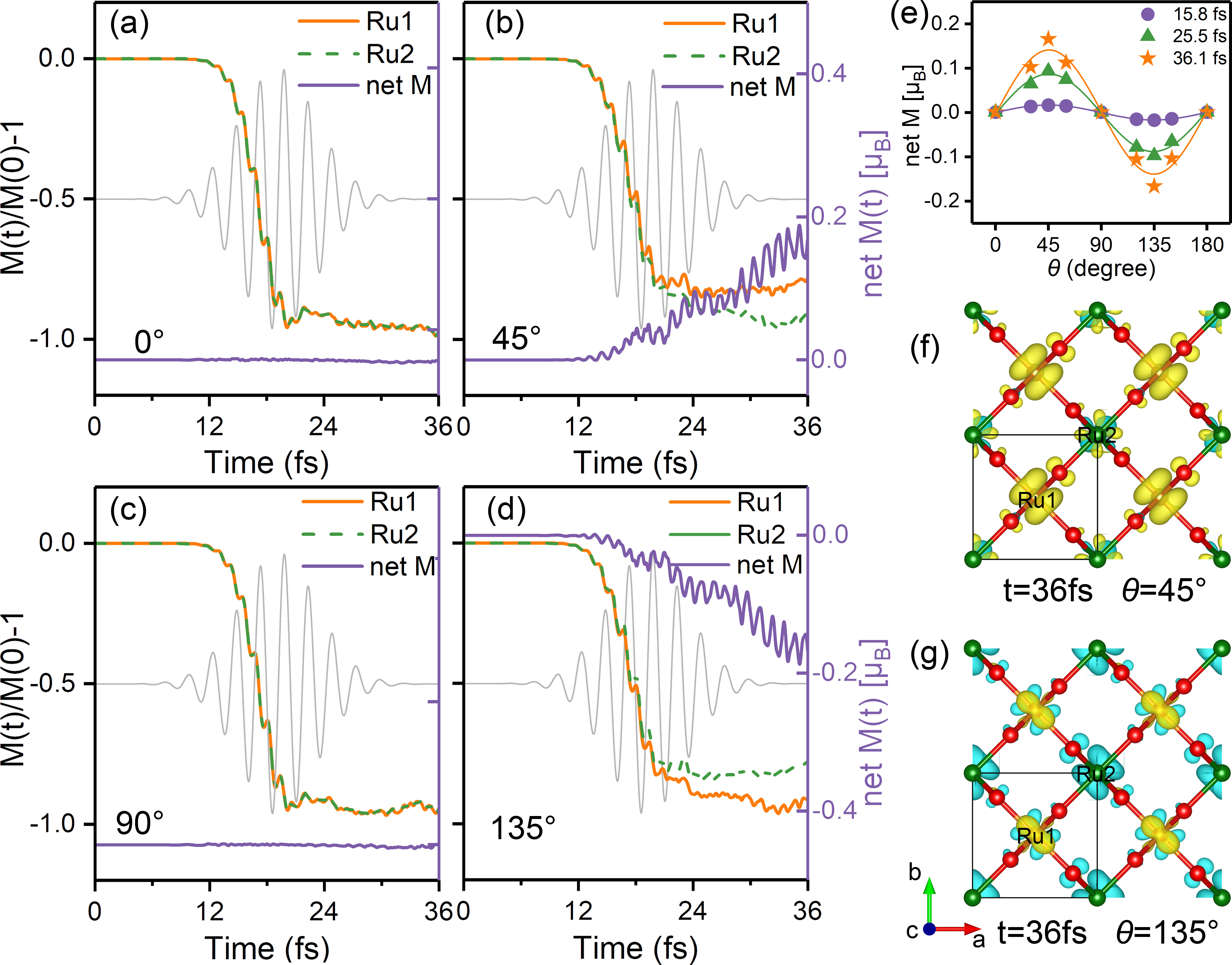}  
  \caption{Polarization-dependent magnetization dynamics in RuO$_2$. (a-d) Normalized Ru atom-resolved spin moment as a function of time at polarization angle $\theta = 0^\circ$, 45$^\circ$, 90$^\circ$ and 135$^\circ$, respectively. The vector potential of the laser pulse is shown in gray (central frequency 1.63 eV, full width at half maximum (FWHM) of $\sim$10 fs, and an incident fluence of 9.8 mJ/cm$^2$). The net magnetic moment (net M) is shown in purple. Corresponding relationship between the net M and $\theta$ at t=15.8 fs, 25.5 fs and 36.1 fs, respectively, is shown in (e). (f) and (g) Magnetization density of RuO$_2$ at $\theta = 45^\circ$ and 135$^\circ$, respectively after the laser dissipates (t=36 fs). Yellow and Cyan domains indicate the spin-up and spin-down density, respectively. The isosurface is set to 0.0045 e/Å$^3$.}
  \label{fig:2}
\end{figure*}

\begin{equation}
\begin{aligned}
i \frac{\partial \psi_j(\mathbf{r},t)}{\partial t}
&=
\Bigg[
\frac{1}{2}\Big(-i\nabla + \frac{1}{c}\mathbf{A}(t)\Big)^2
+ v_s(\mathbf{r},t) \notag \\
&\quad 
+ \frac{1}{2c}\,\boldsymbol{\sigma}\cdot\mathbf{B}_s(\mathbf{r},t) \\
&\quad
+ \frac{1}{4c^2}\,\boldsymbol{\sigma}\cdot\big(\nabla v_s(\mathbf{r},t)\times(-i\nabla)\big)
\Bigg]\psi_j(\mathbf{r},t)\label{1}
\end{aligned}
\tag{1}
\end{equation}

where $\boldsymbol{\sigma}$ represents the Pauli matrices. $v_s(\mathbf{r},t) = v_{\mathrm{ext}}(\mathbf{r},t)
+ v_H(\mathbf{r},t)
+ v_{xc}(\mathbf{r},t)$ is the KS effective potential contributed by the external potential $v_{\mathrm{ext}}$, the classical Hartree potential $v_H$ and the exchange-correlation (XC) potential $v_{xc}$, respectively.  $\mathbf{B}_s(\mathbf{r},t) = \mathbf{B}_{\mathrm{ext}}(\mathbf{r},t) + \mathbf{B}_{xc}(\mathbf{r},t)$ is the KS magnetic field with the magnetic field of the applied laser pulse plus an additional magnetic field $\mathbf{B}_{\mathrm{ext}}$ and XC magnetic field $\mathbf{B}_{xc}$, respectively. The last term in Eq. (1) stands for spin-orbit coupling (SOC) term.

The interaction with the laser field is included using the vector potential:
\begin{equation}
\begin{aligned}
\mathbf{A}(t) =c\frac{E_0}{\omega}\, f(t)\, \sin(\omega t + \phi)\, \hat{\mathbf{e}}_{\theta}\label{2}
\end{aligned}
\tag{2}
\end{equation}

where ${E_0}$ is the electric field amplitude, $\omega$ is the laser field frequency, $c$ is the speed of light, $f(t)$ is the pulse envelope, $\phi$ is the phase, $\hat{\mathbf{e}}_{\theta}$ is a linearly polarized unit vector pointing to angle $\theta$ measured in the $xy$ plane above the $x$ axis [Fig. 1].

All calculations were performed by rt-TDDFT as implemented through the full-potential augmented plane-wave ELK code \cite{20} with a time step of $\Delta$t=2.4 attoseconds. All calculations adhered to the adiabatic local spin density approximation (ALSDA), consistent with methodologies established in our previous works \cite{zhou1,zhou,He1,zhou2,sangeeta1,sangeeta2,sangeeta3,sangeeta4}. More details on the computational methodology can be found in the Supplemental Material \cite{x1}. Due to computational limits, our simulations are restricted to a single unit cell, thus cannot capture spatially extended processes, such as magnon, electron–phonon scattering, and superdiffusive spin currents.

\textit{General consideration for a-OISTR and a-SF}---Conventional \textit{s}-wave magnets (both FMs and AFMs), with their nodal-less Fermi surfaces, enforce fully symmetric ultrafast dynamics: OISTR and SOC-mediated SF act identically on all magnetic sublattices, yielding completely symmetric demagnetization under laser excitation\cite{10}.

However, AMs with spin nodal structures behave in a very different way. As shown in Fig.~S1 in \cite{x1}, the $d$-wave AM RuO$_2$ hosts two vertical spin-degenerate nodal planes. Our band-structure calculations reveal a pronounced anisotropy with strongly momentum (\textbf{\textit{k}})-dependent spin splitting of 1.1~eV near the Fermi level [Fig.~S1(d)], appearing along the M--$\Gamma$--M$^{\prime}$ and S--$\Gamma$--S$^{\prime}$ directions, while X--$\Gamma$--X$^{\prime}$ and Y--$\Gamma$--Y$^{\prime}$ directions remain spin-degenerate.  
This \textbf{\textit{k}}-resolved spin texture produces both symmetric and asymmetric local density of states (LDOS) across the two Ru sublattices, as shown in Figs.~1(d)-1(f). Specifically, Ru2 exhibits spin-up polarization along M--$\Gamma$--M$^{\prime}$ path, whereas Ru1 exhibits spin-down polarization along S--$\Gamma$--S$^{\prime}$ path, yielding a clear LDOS asymmetry; by contrast, nodal direction including X--$\Gamma$--X$^{\prime}$ and Y--$\Gamma$--Y$^{\prime}$, exhibit symmetric LDOS, analogous to AFM behavior.

This broken sublattice symmetry enable an \emph{asymmetrical} (a) OISTR as well as previously unknown spin-flip (a-SF). When the laser polarization $\hat{\mathbf{e}}_{\theta}$ aligns with a spin-splitting direction, e.g. M--$\Gamma$--M$^{\prime}$ or S--$\Gamma$--S$^{\prime}$ path, the photoexcited spin transfer becomes unbalanced, driving a sublattice-selective spin redistribution that breaks the symmetric AFM-like demagnetization [Figs.~1(d) and 1(e)]. In contrast, $\hat{\mathbf{e}}_{\theta}$ along the nodal (spin-degenerate) direction (X--$\Gamma$--X$^{\prime}$ or Y--$\Gamma$--Y$^{\prime}$) yields fully symmetric OISTR.

After the a-OISTR process, the system enters a transient state with unequal charge redistribution between the two sublattices, which subsequently drives an a-SF process near the Fermi level. The LDOS of Ru1 and Ru2 shows sign-reversed features along the M--$\Gamma$--M$'$ and S--$\Gamma$--S$'$ directions, making the laser-driven SF intrinsically unequal [Fig.~1]. For $\hat{\mathbf{e}}_{\theta}$ along M--$\Gamma$--M$'$, SF transitions are dominated by Ru2; along S--$\Gamma$--S$'$ the dominance reverses due to the opposite spin splitting. By contrast, when the $\hat{\mathbf{e}}_{\theta}$ lies along the spin-degenerate nodal planes X--$\Gamma$--X$'$ or Y--$\Gamma$--Y$'$, the LDOS of the two sublattices becomes identical and the resulting spin dynamics revert to AFM-like symmetric behavior. 

These ground-state spin textures in AMs establish the conditions for the emergence of a-OISTR and a-SF effects. To quantify how these asymmetries evolve on femtosecond timescales, we next employ rt-TDDFT to simulate the resulting ultrafast magnetization response.

\textit{Polarization-dependent magnetization}--- We demonstrate this physics by driving RuO$_2$ with linearly polarized laser pulses oriented along different $\hat{\mathbf{e}}_{\theta}$. Under laser excitation, the Ru atoms (initial spin moment 1.19~$\mu_B$) undergo ultrafast demagnetization. Figures~2(a)-2(d) show the normalized spin moments of Ru1 and Ru2 as a function of time for four polarization angles $\theta$.
When the $\hat{\mathbf{e}}_{\theta}$ is parallel to the spin-degenerate nodal planes ($\theta = 0^\circ$ and $90^\circ$), the local spin moment of Ru1 and Ru2 ($M_1$ and $M_2$) lose equally, preserving symmetric demagnetization. In contrast, for $\theta = 45^\circ$ and $135^\circ$—i.e., $\hat{\mathbf{e}}_{\theta}$ along the spin-polarized planes— Ru1 and Ru2 demagnetize unequally, generating a transient ferrimagnetic state with a net moment ($M_{net}=M_1 - M_2$) of $\sim$0.2~$\mu_B$ within 36~fs. Notably, Ru2 demagnetizes more strongly than Ru1 at $\theta=45^\circ$, whereas the sign reverses at $\theta=135^\circ$. Further demagnetization dynamics within 120 fs indicate that this ferrimagnetic state can be sustained for more than 100 fs, demonstrating that the laser-induced magnetization remains robust and persists for a considerable duration [Fig. S2].

\begin{figure}
  \includegraphics[width=0.5\textwidth]{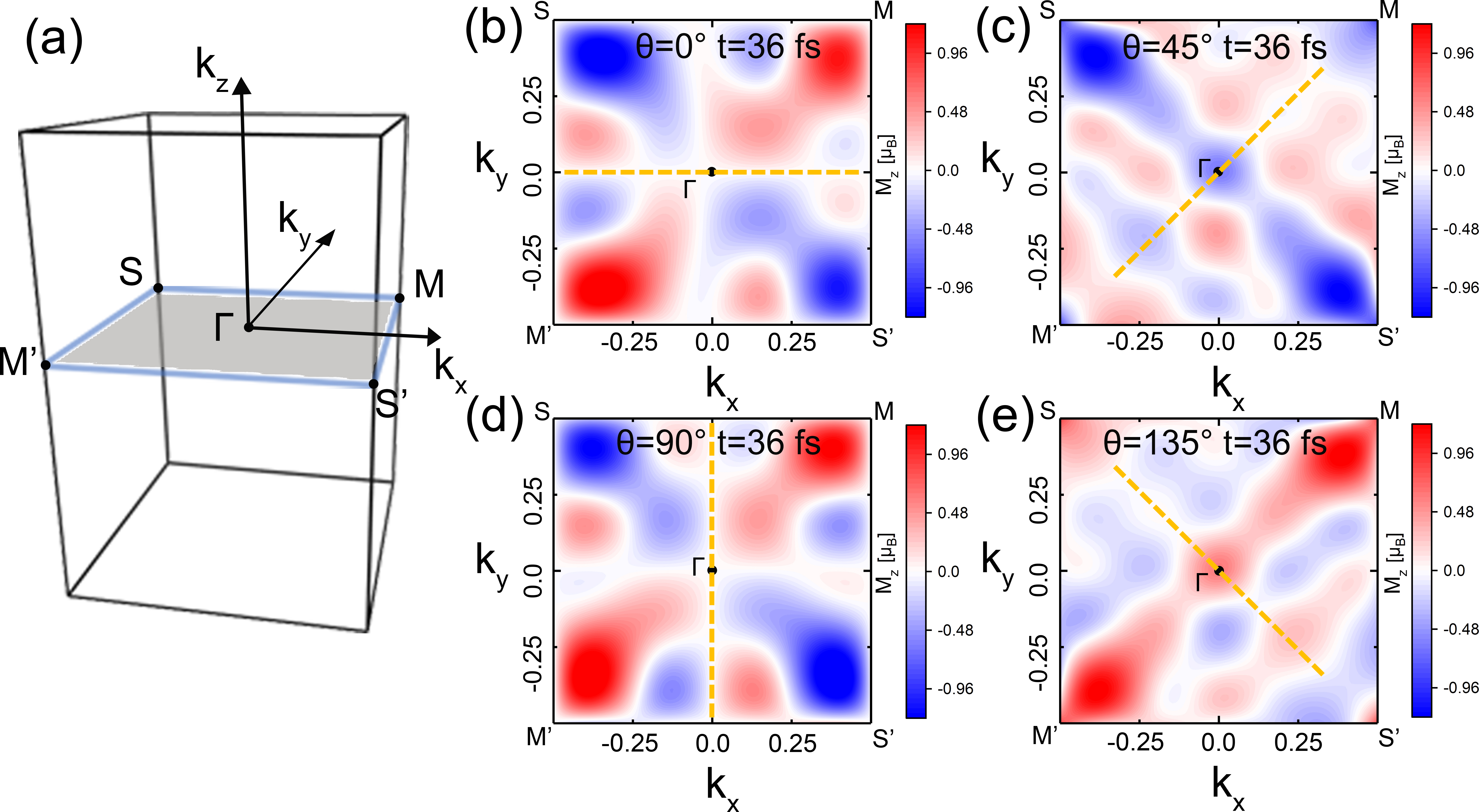}  
  \caption{Polarization-dependent transient magnetization distribution in \textbf{\textit{k}}-space. (a) 3D Brillouin zone of RuO$_2$. (b-e) Snapshots of magnetization distribution at $k_z$=0 and t=36 fs with $\theta = 0^\circ$, 45$^\circ$, 90$^\circ$ and 135$^\circ$, respectively. Color intensity indicates magnetization magnitude; Yellow dashed lines represent the $\hat{\mathbf{e}}_{\theta}$ direction.}
  \label{fig:3}
\end{figure}

\begin{figure*}
  \includegraphics[width=0.95\textwidth]{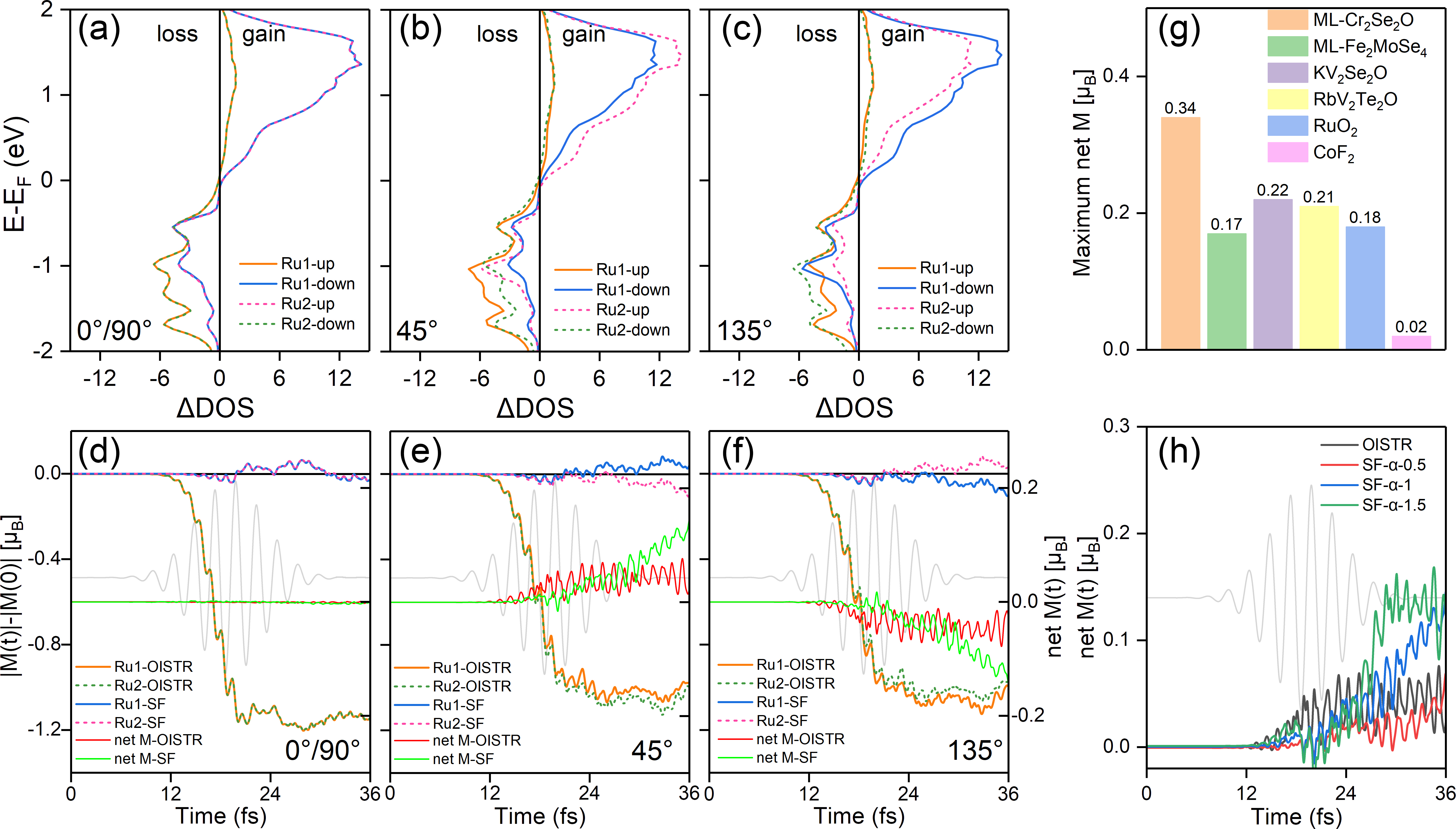}  
  \caption{Contribution of OISTR and SF process ($M_{OISTR}$ and $M_{SF}$) on demagnetization of Ru atoms in RuO$_2$. (a-c) Differences in the time-resolved occupation function $\Delta$DOS(t) in the absence of SOC at $\theta = 0^\circ/90^\circ$, $\theta = 45^\circ$ and $\theta = 135^\circ$. The negative value signifies a loss of electrons, and a positive value signifies a gain of electrons. (d-f) The time-dependent $M_{OISTR}$ and $M_{SF}$ for two Ru atoms and change in net M at $\theta = 0^\circ/90^\circ$, $\theta = 45^\circ$ and $\theta = 135^\circ$, respectively. (g) Laser-induced maximum net moment ($M_{net}$) of six \textit{d}-wave AMs including monolayer (ML) Cr$_2$Se$_2$O, ML-Fe$_2$MoSe$_4$, KV$_2$Se$_2$O, RbV$_2$Te$_2$O, RuO$_2$ and CoF$_2$ during the initial 36 fs. (h) Time evolution of the $M_{OISTR}$ and $M_{SF}$ of RuO$_2$ with different SOC strengths ($\alpha_{SO}$).}
  \label{fig:4}
\end{figure*}

Figure 2(e) displays the $M_{net}$ as a function of the $\theta$ for three representative transient times. At all times, the $M_{net}$ follows a clear sinusoidal dependence, $M_{\mathrm{net}} \propto \sin\theta$,
 with a 180$^\circ$ periodicity. This twofold symmetry directly reflects the interplay between the linearly polarized field and the underlying \textit{d}-wave nodal spin structure. In particular, the 180$^\circ$ oscillation encodes the orientation of the spin-degenerate nodal planes and the \textbf{\textit{k}}-dependent spin splitting inherent to the \textit{d}-wave spin texture. Thus, the sinusoidal angular response establishes a direct link between the magnetization dynamics, $\hat{\mathbf{e}}_{\theta}$, and the \textit{d}-wave electronic structure.

To further unveil asymmetric demagnetization dynamics, we visualize the magnetization density in real-space, as shown in Figs. 2(f), 2(g) and Fig. S3, alongside the transient magnetization distribution at the $k_z$=0 plane in \textbf{\textit{k}}-space [Fig. 3] after the laser pulse (t=36 fs). Figure S3 shows that the ground state magnetization densities (before laser) of Ru1 and Ru2 are equal, consistent with previous work \cite{24}. Upon laser excitation, a striking asymmetry emerges. At $\theta = 45^\circ$, the magnetization density of Ru1 sublattice surpasses (falls below) that of Ru2, while at $\theta = 135^\circ$, the opposite occurs. In contrast, at $\theta = 0^\circ$ and 90$^\circ$, the magnetization densities remain equivalent [Fig. S3]. Such anisotropic magnetization densities result in the symmetric (asymmetric) demagnetization observed along the $\hat{\mathbf{e}}_{\theta}$ parallel to the spin-degenerate nodal planes (spin-polarized planes). This asymmetry is further reflected in \textbf{\textit{k}}-space. At $\theta = 0^\circ$ and 90$^\circ$, the excited magnetization distribution in \textbf{\textit{k}}-space exhibits alternating and symmetric patterns [Figs. 3(b) and 3(d)]. However, at $\theta = 45^\circ$ (135$^\circ$), this alternation of excited magnetization is disrupted--- magnetization along the M-$\Gamma$-M$^{\prime}$ (S-$\Gamma$-S$^{\prime}$) path becomes smaller compared to the S-$\Gamma$-S$^{\prime}$ (M-$\Gamma$-M$^{\prime}$) path [Figs. 3(c) and 3(e)]. The remarkable agreement between the magnetization density in real space and the transient magnetization distribution in \textbf{\textit{k}}-space highlights the central role of the a-OISTR mechanism in driving asymmetric demagnetization in AMs.

The $\hat{\mathbf{e}}_{\theta}$-dependent spin-resolved charge transfer further supports this picture [Fig.~S4]. Although both Ru atoms lose spin moment, an obvious asymmetry emerges when the $\hat{\mathbf{e}}_{\theta}$ aligns with the spin-polarized planes. At $\theta = 45^\circ$ ($135^\circ$), spin-resolved occupation changes are substantially larger on Ru2 (Ru1), indicating a $\hat{\mathbf{e}}_{\theta}$-selective flow of spin-dependent charge. This behavior directly agree with the $\hat{\mathbf{e}}_{\theta}$-dependent asymmetric demagnetization.

\textit{Demagnetization via a-OISTR and a-SF}--- Within the tens-of-femtoseconds window, OISTR drives the earliest field-induced asymmetric spin transfer and triggers $M_{net}$, while SOC-mediated SF emerges shortly thereafter and amplifies it \cite{krieger2015laser}. Thus, both processes coexist on this timescale and act together to produce the ultrafast demagnetization \cite{ryan2023optically}. To disentangle their contributions, we switch off SOC and analyze the time-resolved change in the DOS, $\Delta\mathrm{DOS}(t)$, between $t = 25.3$~fs and $0$~fs [Figs.~4(a)–4(c)], isolating the a-OISTR response. Even without SOC, a strong $\hat{\mathbf{e}}_{\theta}$-selective asymmetry remains: electrons preferentially occupy the Ru2 spin-up and Ru1 spin-down channels, generating a spin-resolved inter-sublattice current and reducing $M_1$ and $M_2$. Note that this response is strongly polarization-selective. At $\theta=45^\circ$ ($135^\circ$), the enhanced Ru2 (Ru1) occupation reflects the opposite spin splitting along the M--$\Gamma$--M$^{\prime}$ and S--$\Gamma$--S$^{\prime}$ directions and correlates with the asymmetric demagnetization. In contrast, at $\theta=0^\circ$ and $90^\circ$, the charge transfer becomes nearly symmetric, consistent with symmetric demagnetization. This $\hat{\mathbf{e}}_{\theta}$-dependent evolution of TDDOS is uniquely fingerprints a-OISTR in \textit{d}-wave RuO$_2$.

The SF transitions, unlike OISTR, do not conserve total magnetization and thus amplify the asymmetric demagnetization between Ru1 and Ru2 [Figs.~2(b) and 2(d)], enhancing $M_{net}$. We decompose $M_{net}$ into the OISTR and SF components ($M_{net} = M_{OISTR} + M_{SF}$) under different polarizations in Figs.~4(d)–4(f). The $M_{OISTR}$ is dominated in the 12–24~fs window, producing a rapid loss of $\sim1 \mu_\mathrm{B}$ for $M_1$ and $M_2$. After $\sim$24~fs, $M_{OISTR}$ saturates as the pulse ends. $M_{SF}$, initially small, grows gradually after the pulse and becomes significant during 24–36~fs. $M_{net}$ follows the same trend: $M_{OISTR}$ drives the early-time rise, while $M_{SF}$ overtakes it around 24–36~fs and governs the late-time evolution.

Both $M_{OISTR}$ and $M_{SF}$ exhibit strong $\theta$-dependence, consistent with the results in Fig.~1. At $\theta=0^\circ$/$90^\circ$, $M_{OISTR}$ and $M_{SF}$ remain symmetric between Ru1 and Ru2, yielding $M_{net}=0$. At $\theta=45^\circ$, both become asymmetric. Importantly, the $M_{SF}$ of Ru1 and Ru2 remain nearly symmetric in the early stage (12–20~fs), but diverge after $\sim20$~fs: $M_{SF} \textgreater 0$ for Ru1 while $M_{SF} \textless 0$ for Ru2. This sign, which is opposite to the $M_{OISTR}$ for Ru1 and Ru2, further enhances their imbalance in the demagnetization, resulting in a large positive magnetization, $M_{net} \textgreater 0$. At $\theta=135^\circ$, the signs of both $M_{OISTR}$ and $M_{SF}$ reverse, generating a negative magnetization, $M_{net}  \textless 0$.

These results indicate that a-OISTR and a-SF operate on distinct timescales: OISTR governs the ultrafast, laser-driven asymmetric demagnetization in the first tens of femtoseconds, whereas SF becomes the primary source of asymmetric dynamics after the pulse fades, continuously reinforcing the $M_{net}$. The a-OISTR sets the initial asymmetrical magnetization dynamics by generating a strongly $\theta$-dependent spin-polarized charge transfer, while a-SF acts as a secondary mechanism that amplifies this asymmetry after the laser pulse. Although the SF-induced changes in spin moment are an order of magnitude smaller than those from OISTR, their asymmetric character makes them highly effective in shaping the final $M_{net}$.

\textit{Discussion}--- We further examine the influence of $\alpha_{SO}$ (SOC strength) on the demagnetization dynamics, as shown in Fig.~4(h). Increasing the $\alpha_{SO}$ to 1.5 (for heavier elements case) enhances the $M_{SF}$, whereas reducing $\alpha_{SO}$ to 0.5 markedly suppresses the SF contribution. Because Ru exhibits stronger SOC than typical 3$d$ elements, we additionally investigate the recently experimentally confirmed $d$-wave AM KV$_2$Se$_2$O \cite{zhang2025crystal}, [Figs. S5 and S6]. Here, the SOC-induced contribution to the $M_{net}$ is negligible, and the laser-induced magnetization originates predominantly from the OISTR mechanism. Remarkably, even in the absence of SOC, OISTR alone in KV$_2$Se$_2$O is sufficient to generate a strong magnetization, $M_{net} \approx 0.2 \mu_B$ [Fig. S5], demonstrating that OISTR constitutes the primary microscopic mechanism for laser-induced magnetization in $d$-wave AMs.

In addition, we also examine several additional $d$-wave AMs---including experimentally confirmed RbV$_2$Te$_2$O \cite{jiang2025metallic}, as well as theoretically proposed CoF$_2$ \cite{adamantopoulos2024spin}, Cr$_2$Se$_2$O \cite{khan2025altermagnetism}, and Fe$_2$MoSe$_4$ \cite{li2025ferrovalley}, as shown in Figs.~4(g) and ~S6. All systems exhibit similar asymmetric spin dynamics with large net magnetization, ranging from $0.02\,\mu_B$ to $0.34\,\mu_B$, demonstrating that the laser-induced metastable magnetization is generic across $d$-wave AMs. This universality directly reflects the nodal spin texture inherent to their band structures. 

To further study the robustness of our results, we examine the impact of laser pulse parameters (frequency and fluence), local Coulomb interaction and strain on the $M_1$ and $M_2$ loss for RuO$_2$ [Figs. S7–S10]. The same underlying physics consistently emerges. Higher-frequency pulses more efficiently drive interband excitations, yielding a stronger asymmetric demagnetization. While the $M_{net}$ initially increases with fluence, it eventually saturates at a critical value set by the intrinsic band splitting of the AMs. For comparison, we also investigate the spin dynamics of conventional AFM, NiO. We observe that, as expected and unlike RuO$_2$, the demagnetization is totally symmetric for NiO with $M_{net}=0$ [Figs. S11 and S12].

AMs with the presence of SOC, may exhibit a weak ferromagnetism (WFM), originating from crystal symmetry governed orbital angular momentum (OAM, \textbf{L}) anisotropy \cite{Daegeun}. However, the SF process requires non-conservation of the spin projection, $[H,S_z]\neq 0$. While OAM  anisotropy described by the crystal-field (CF) Hamiltonian $H_{\rm CF}=V(\mathbf r)$ acts only on spatial coordinates and commutes with spin, $[H_{\rm CF},\mathbf S]=0$, and thus cannot directly induce SF. However, OAM anisotropy can only indirectly modulate SF amplitudes via SOC matrix elements, $\langle \psi_i|\mathbf L\! \cdot\! \mathbf S|\psi_j\rangle$, which are associated with magnetic anisotropy energies (MAE) \cite{xiang2013magnetic} in the $\mu{\rm eV}\!-\!{\rm meV}$ range. According to $\Delta t\sim\hbar/\Delta E$, this corresponds to picosecond or longer timescales, far slower than those of OISTR and SF. 

Our theoretical predictions can be experimentally validated using ultrafast spectroscopy techniques such as time-resolved magneto-optical Kerr effect (TR-MOKE) and magnetic circular dichroism (MCD). For instance, earlier time-resolved circular-dichroism experiments reported a laser-induced ferrimagnetic state in CoF$_2$ at $\theta=\pm45^\circ$~\cite{disa}, accompanied by a pronounced polarization dependence. Recently, TR-MOKE and MCD have been employed to explore spin dynamics in RuO$_2$ and MnTe on femtosecond timescales \cite{27,28}. Notably, RuO$_2$ exhibits $\theta$-dependent Kerr signals, with a striking reversal of Kerr rotation at $\theta = 45^\circ$ and 135$^\circ$, consistent with present theoretical predictions.  In addition, while our simulations address early-time dynamics (up to 100 fs), the impact of electron–phonon coupling at longer timescales and its possible role in enhancing $M_{\mathrm{net}}$ or driving a transition toward the FM state remains an open question for future work.

\textit{Conclusion}--- Using real-time ab-initio simulations, we have discovered that linearly polarized femtosecond laser pulses can induce a strong magnetization in a compensated \textit{d}-wave AMs. We identified the microscopic origin of this phenomenon and established a two-step mechanism, a-OISTR and a-SF, that governs the resulting ultrafast magnetization dynamics. Specifically: (i) Laser generates a robust metastable ferrimagnetic state in RuO$_2$ with a $M_{net}\approx 0.2 \mu_B$ per unit cell, whose sign and magnitude are fully controlled by the light polarization. (ii) a-OISTR arising from the alternating spin splitting, creates the net magnetization during the pulse by driving a strongly direction-selective spin-polarized inter-sublattice current. This mechanism is unique to \textit{d}-wave AMs and cannot occur in conventional magnets. (iii) Once this sublattice imbalance is established via a-OISTR, a previously unknown a-SF mechanism becomes active. This a-SF process does not create magnetization during the laser excitation, but amplifies and stabilizes it after the laser pulse. (iv) The light–matter mechanism revealed in RuO$_2$ is fundamentally the same in other \textit{d}-wave AMs, such as KV$_2$Se$_2$O and RbV$_2$Te$_2$O. Our results not only provide a fundamental understanding of laser-induced ultrafast magnetization dynamics in AMs, but also lay the groundwork for future developments in ultrafast altermagnetic spintronics, where a phase transition of AMs to ferrimagnetic state can be used as a key ingredient. 

\textit{Acknowledgments}---Z. Z would like to thank the support from MSCA Fellowships CZ-UK3 $\text{CZ.02.01.01/00/22\_010/0008820}$. S.S would like to thank the Leibniz Professorin program (SAW P118/2021) for support. J. K. D and S. S would like to thank DFG for funding through project-ID 328545488 TRR227 (project A04). We thank the e-INFRA CZ (ID:90140) for providing computational resources.

\bibliographystyle{apsrev4-2}
\bibliography{apssamp}

\end{document}